\documentstyle[12pt,twoside,epsfig]{article}
\begin{document}
\title{{\bf Quantum Scalar-metric Cosmology with Chaplygin gas}}
\author{Barun Majumder\thanks{Email: barunbasanta@iiserkol.ac.in}\\\\
{\small {\it Department of Physical Sciences,}}\\{\small {\it Indian Institute of Science Education and Research (Kolkata),}}\\{\small {\it India}}}
\maketitle
\begin{abstract} 
\hspace{-5pt} A spatially flat Friedmann-Robertson-Walker(FRW) cosmological model with generalized Chaplygin gas is studied in the context of scalar-metric formulation of cosmology. Schutz's mechanism for the perfect fluid is applied with generalized Chaplygin gas and the classical and quantum dynamics for this model is studied. It is found that the only surviving matter degree of freedom played the role of cosmic time. For the quantum mechanical description it is possible to find the wave packet which resulted from the linear superposition of the wave functions of the Schr\"{o}dinger-Wheeler-DeWitt(SWD) equation, which is a consequence of the above formalism. The wave packets show two distinct dominant peaks and propagate in the direction of increasing scale factor. It may happen that our present universe originated from one of those peaks. The many-world and ontological interpretation of quantum mechanics is applied to investigate about the behaviour of the scale factor and the scalar field(considered for this model). In both the cases the scale factor avoids singularity and a bouncing non-singular universe is found.
\vspace{5mm}\newline Keywords: quantum scalar-metric cosmology, S\'aez-Ballester theory, Chaplygin gas, Schutz formalism
\end{abstract}

\section{Introduction}

In recent years an exotic fluid known as Chaplygin gas has well described the reason behind the current observation of cosmic acceleration. It was first proposed in \cite{r1}. The generalized Chaplygin gas satisfies an equation of state $p=-\frac{A}{\rho^{\alpha}}$, where $p$ is the pressure, $A$ is a positive constant and $0<\alpha \leq 1$ \cite{r1,r2}. $\alpha=1$ corresponds to standard Chaplygin gas \cite{r1}. This equation of state describes an universe evolving from a non-relativistic matter dominated one to a cosmological constant dominated one \cite{r2}. The equation of state for the Chaplygin gas actually comes from the string Nambu-Goto action in the light-cone coordinate \cite{r3}. The idea of this exotic fluid was later used widely \cite{r3,r4,r5,r6,r7,r8,r9,r10} mainly to find a solution to the coincidence problem in cosmology. The isotropic quantum cosmological models with Chaplygin gas have been studied very recently \cite{r11,r12}. The generalized Chaplygin gas is also studied in a modified gravity approach \cite{r13} and the geometrical explanation for the generalized Chaplygin gas gives the deceleration parameter and the age of the universe which are consistent with observation \cite{r14}.
\par
Here we shall study the quantum dynamics of a spatially flat FRW model with Chaplygin gas and a dimensionless scalar field coupled to the metric. This formulation was first classically studied by S\'aez and Ballester \cite{r15} and there a satisfactory answer to the missing-matter problem in cosmology is given. Recently the quantum description of this S\'aez-Ballester theory in a generalized form is studied \cite{r16,r17,r18}. In our case we take a polynomial coupling of the form $6\lambda^2\phi^m$ to the metric. We then apply Schutz's formalism of a perfect fluid \cite{r19,r20}, which was extensively used in some recent publications \cite{r21,r22,r23,r18} to the Chaplygin gas. With the canonical methods \cite{r24} we get the super Hamiltonian. We studied the classical dynamics of the system and found that the remaining matter degree of freedom can play the role of time. Canonical quantization of the super Hamiltonian gives us the Schr\"odinger-Wheeler-DeWitt(SWDW) equation. Then we linearly superpose the wave functions of the SWDW equation to construct the wave packet. In the next step we recovered the classical solutions using the many-world and ontological interpretation of quantum mechanics.

\section{The Quantum Model}

The expression for action in this spatially flat quantum FRW model with generalized Chaplygin gas can be written as
\begin{equation}
{\cal A} = \int_Md^4x\sqrt{-g}\,\left[R-F(\phi)g^{\mu \nu}\phi_{,\mu}\phi_{,\nu}\right] + 2\int_{\partial 
M}d^3x\sqrt{h}\, h_{ab}\, K^{ab} + \int_Md^4x\sqrt{-g}\,\, P_c
\end{equation}
where $h_{ab}$ is the induced metric over three dimensional spatial hypersurface which is the boundary $\partial M$ of the four dimensional manifold M and $K^{ab}$ is the extrinsic curvature.Here units are so chosen that $c=16\pi G=\hbar$ is equal to one.$P_c$ is the pressure for generalized Chaplygin gas. Here $F(\phi)$ is an arbitrary function of a scalar field which is coupled to the metric. The first term(with $R$ only) and the third term were first obtained in \cite{r25}. The generalized Chaplygin gas satisfies the equation of state
\begin{equation}
P_c=-\frac{A}{\rho_c ^{\alpha }}.
\end{equation}
Here  $A$ is a positive constant and $0<\alpha \leq 1$. $\alpha =1$ represents the standard Chaplygin gas.In Schutz's formalism \cite{r19,r20} the fluid's four velocity can be expressed in terms of three potentials $\epsilon$, $\theta$ and $S$ (here we are studying spatially flat FRW model so other potentials are absent in this model because of its symmetry),
\begin{equation}
u_\nu = \frac{1}{h}(\epsilon_{,\nu} + \theta S_{,\nu}).
\end{equation}
Here $h$ is the specific enthalpy, $S$ is the specific entropy, $\epsilon$ and $\theta$ have no direct physical
meaning.The four velocity also satisfy the normalization condition
\begin{equation}
u^\nu u_\nu = 1.
\end{equation}
The metric for the spatially flat FRW model is
\begin{equation}
ds^2=N^2(t)dt^2-a^2(t)\left[{dr^2}+r^2(d\vartheta^2+ \sin^2 \vartheta d\varphi^2)\right],
\end{equation}
where $N(t)$ is the lapse function and $a(t)$ the scale factor.
If we study the thermodynamics of a perfect fluid \cite{r24} we can see that $h$ and $S$ satisfy these thermodynamic relations
\begin{equation}
\label{thermo}
\rho = \rho_0(1+\Pi)\quad ,\quad h = 1+\Pi+P/\rho_0 \quad ,\quad \tau dS = d\Pi+P d(1/\rho_0) \quad
\end{equation}
where $\tau$ and $\rho$ are temperature and particle number density respectively.We can use Schutz's mechanism with these thermodynamic relations to evaluate the third term of the action.
If we eliminate the surface terms , the first and second term of the action gives
\begin{equation}
{\cal A}_g = \int dt\biggr[-\frac{6}{N}a\dot{a}^2+\frac{1}{N}F(\phi)a^3\dot{\phi}^2\biggl] .
\end{equation}
So the gravitational Lagrangian density is
\begin{equation}
{\cal L}_g = -\frac{6}{N}a\dot{a}^2+\frac{1}{N}F(\phi)a^3\dot{\phi}^2.
\end{equation}
The conjugate momenta for these $a$ and $\phi$ are then
\begin{equation}
\label{aphi}
p_a = - 12\frac{a}{N}\dot a \quad , \quad p_{\phi} = 2\frac{a^3}{N}F(\phi)\dot {\phi} ,
\end{equation}
and the Hamiltonian for this part is
\begin{equation}
{\cal H}_g = N\left[-\frac{1}{24}\frac{P_a^2}{a}+\frac{1}{4F(\phi)a^3}P_{\phi}^2\right].
\end{equation}
Let us now find the Hamiltonian for the generalized Chaplygin gas.Rewriting the third equation of (\ref {thermo})
\begin{equation}
\tau dS =\frac{(1+\Pi)^{-\alpha}}{1+\alpha}d\left[(1+\Pi)^{1+\alpha}+\frac{A}{\rho_0^{1+\alpha}}\right].
\end{equation}
It then follows to within a factor $S=(1+\Pi)^{1+\alpha}+\frac{A}{\rho_0^{1+\alpha}}$ and $\tau=\frac{(1+\Pi)^{-\alpha}}{1+\alpha}$. The equation of state $P_c=-\frac{A}{\rho_c ^{\alpha }}$ takes the form 
\begin{equation}
P_c=-A\left[\frac{1}{A}\left(1-\frac{\,\,h^{\frac{1+\alpha}{\alpha}}}{S^{1/\alpha}}\right)\right]^{\frac{1+\alpha}{\alpha}}.
\end{equation}
If we consider that the four velocity has only one component in this model $u_\nu=(N,0,0,0)$ then with the normalization condition we find that
\begin{eqnarray}
\label{ac}
{\cal A}_c = \int dt\biggr\{-N a^{3}
A\left[\frac{1}{A}\left(1-\frac{\,\,(\dot{\epsilon}+\theta\dot{S})^{\frac{1+\alpha}{\alpha}}}{N^{\frac{1+\alpha}{\alpha}}
S^{1/\alpha}}\right)\right]^{\frac{1+\alpha}{\alpha}}\biggr\}.
\end{eqnarray}
As $h>0$ so $(\dot\epsilon + \theta\dot S)>0$. If we try the canonical methods \cite{r24} to find the Hamiltonian for this action we will end up with 
\begin{equation}
{\cal H}_c =N \left(S p_{\epsilon}^{1+\alpha}+A a^{3(1+\alpha)} \right)^{\frac{1}{1+\alpha}},
\end{equation}
where $p_\epsilon =\frac{\displaystyle \partial{\cal L}_c}{\displaystyle \partial \dot{\epsilon}}$ and ${\cal L}_c$ is the expression inside second bracket of equation (\ref {ac}).
Let us now write the super Hamiltonian for the minisuperspace of this FRW model and that is
\begin{eqnarray}
\label{sh}
{\cal H}={\cal H}_g+{\cal H}_c = N\left[-\frac{1}{24}\frac{P_a^2}{a}+\frac{1}{4F(\phi)a^3}P_{\phi}^2+\left(S p_{\epsilon}^{1+\alpha}+A a^{3(1+\alpha)} \right)^{\frac{1}{1+\alpha}}\right].
\end{eqnarray}
Analytical solution of the Schr\"odinger-Wheeler-DeWitt equation formed from this super Hamiltonian is very hard to find.So some meaningful approximation can be made for the last term. For the early universe scenario we can assume $S p_{\epsilon}^{1+\alpha}\gg A a^{3(1+\alpha)}$ and using this \cite{r11} we can expand the expression of the last term of equation (\ref {sh}).
\begin{equation}
\hspace{-1cm}\left(S p_{\epsilon}^{1+\alpha}+ A a^{3(1+\alpha)}\right) ^{\frac1{1+\alpha}}\approx S^{\frac{1}{1+\alpha}} p_{\epsilon}\left[1+\frac1{1+\alpha}\frac{ A a^{3(\alpha+1)}}{S p_{\epsilon}^{1+\alpha}}+\frac{-\alpha}{2{(1+\alpha)}^2}\frac{A^{2}}{S^2    p_{\epsilon}^{2(1+\alpha)}}a^{6(\alpha+1)}+\ldots\right]. 
\end{equation}
Up to the leading order, the super-Hamiltonian takes the form
\begin{equation}
{\cal H}=N\left[-\frac{1}{24}\frac{P_a^2}{a}+\frac{1}{4F(\phi)a^3}P_{\phi}^2 + S^{\frac{1}{1+\alpha}} p_{\epsilon}\right].
\end{equation}
The canonical transformation
\begin{eqnarray}
T =-(1+\alpha)p_\epsilon^{-1}  S^{\frac{\alpha}{1+\alpha}}p_S  \quad ,
\quad p_{T} =S^{\frac{1}{1+\alpha}} p_\epsilon
\end{eqnarray}
would further allow us to simplify the super Hamiltonian and finally
\begin{equation}
\label{sh1}
{\cal H}=N\left[-\frac{1}{24}\frac{P_a^2}{a}+\frac{1}{4F(\phi)a^3}P_{\phi}^2 + P_T\right].
\end{equation}
Here $P_T$ is the only canonical variable associated with matter. Now if we can relate any of the canonical variables with time, we can have a well defined Hilbert space structure. Let us now study the classical dynamics governed by the super Hamiltonian. The Hamilton's equation of motion are
\begin{eqnarray}
\label{he}
\left\{
\begin{array}{ll}
\dot{a}=\{a,{\cal H}\}=-\frac{N}{12}\frac{P_a}{a},\\\\
\dot{P_a}=\{P_a,{\cal H}\}=N\left[-\frac{1}{24}\frac{P_a^2}{a^2}+\frac{3}{4F(\phi)a^4}P_{\phi}^2\right],\\\\
\dot{\phi}=\{\phi,{\cal H}\}=\frac{N}{2F(\phi)a^3}P_{\phi},\\\\
\dot{P_{\phi}}=\{P_{\phi},{\cal H}\}=N\frac{P_{\phi}^2}{4a^3}\frac{F'(\phi)}{F(\phi)^2},\\\\
\dot{T}=\{T,{\cal H}\}=N,\\\\
\dot{P_T}=\{P_T,{\cal H}\}=0,
\end{array}
\right.
\end{eqnarray}
with the constraint ${\cal H}=0$. Actually $N$ acts as a Lagrange multiplier for the system. If we see the last two equations of (\ref{he}) we can see that if $N=1$, then $T=t$. So $T$ may play the role of cosmic time if we choose the gauge $N=1$. Also $P_T=C_T$(a constant) from the last equation. So now with $N=1$ and $P_T=C_T$ the third and fourth equation of (\ref{he}) gives
\begin{equation}
\frac{\ddot{\phi}}{\dot{\phi}}+\frac{1}{2}\frac{F'(\phi)}{F(\phi)}\dot{\phi}+3\frac{\dot{a}}{a}=0,\end{equation}
which on integration yields
\begin{equation}
\label{c}
\dot{\phi}^2 F(\phi)=C a^{-6},
\end{equation}
where $C$ is an integration constant. The constraint ${\cal H}=0$ can be written in terms of canonical co-ordinates with the help of equation (\ref{he})
\begin{equation}
\label{ct}
-6a\dot{a}^2+\dot{\phi}^2F(\phi)a^3+C_T=0.
\end{equation}
With equation (\ref{c}) and equation (\ref{ct}) we can solve for the scale factor and it is
\begin{equation}
\label{sf}
a(t)=\left[\frac{3C_T}{8}(t+C_1)^2-\frac{C}{C_T}\right]^{\frac{1}{3}},
\end{equation}
where $C_1$ is an integration constant and we can set it to $C_1=\sqrt{\frac{8C}{3}}\frac{1}{C_T}$ to satisfy $a(t=0)=0$. We will now consider the case $F(\phi)=6\lambda^2\phi^m$, where $\lambda$ and $m$ are constants and $m\neq -2$ (actually $m\ge0$ give us good results for all values of $m$). With this and using equation (\ref{c}) and equation (\ref{sf}) we get
\begin{equation}
\label{fit}
\phi(t)=\left\{\frac{(m+2)}{6\lambda}\ln \left[\frac{C_2t}{t+2C_1}\right]\right\}^{\frac{2}{m+2}},
\end{equation}
\begin{figure}
\label{field}
\begin{tabular}{c}
\includegraphics[width=7cm,height=6cm]{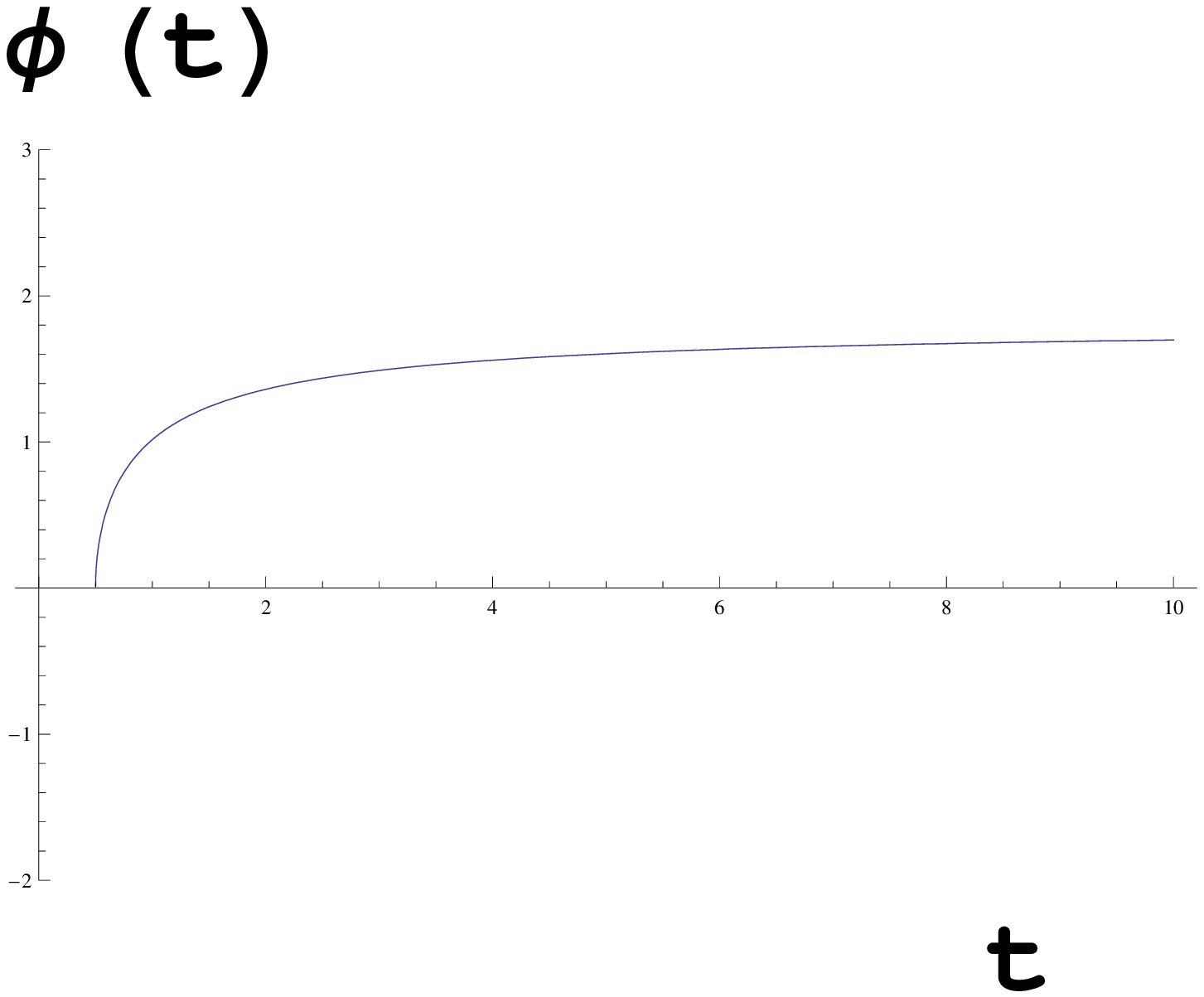} \hspace{1cm}
\includegraphics[width=7cm,height=6cm]{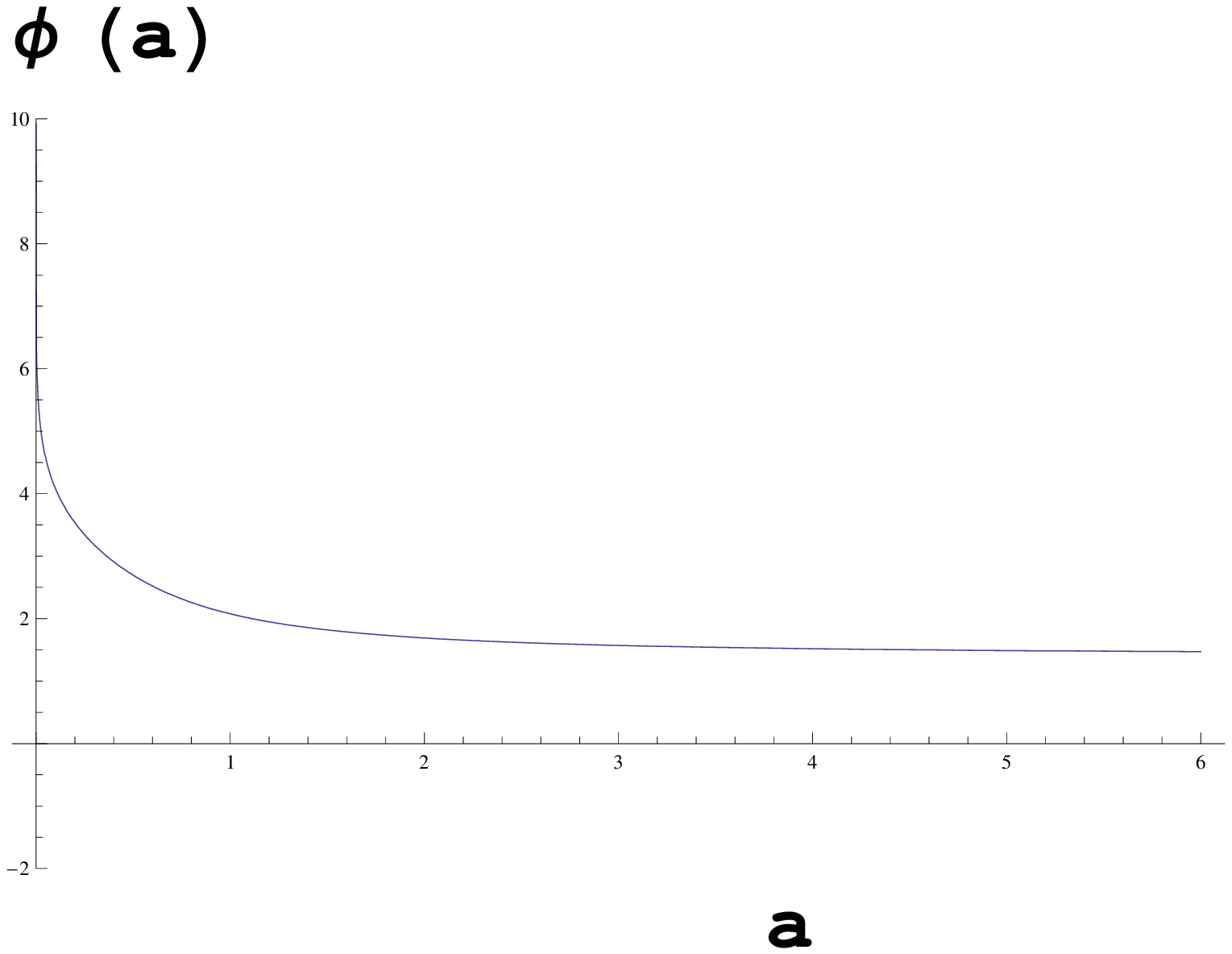}
\end{tabular}
\caption{\footnotesize  Left:The classical behaviour of $\phi$ at late times. Right:The classical behaviour of $\phi$ as a function of the scale factor. The plots are with numerical values $m=2, \lambda=0.33, C=0.375, C_T=1, C_1=1, C_2=5, C_3=2$. With other admissible values the same nature of the curves are repeated.}
\end{figure}
where $C_2$ is a constant of integration. From equation (\ref{fit}) we can see that $\phi(t=$finite$)\rightarrow constant$.The behaviour of $\phi$ at the beginning is not well understood as shown in figure $1$ (left). Let us now find the expression of $\phi$ as a function of the scale factor. With the help of equation (\ref{c}), equation (\ref{ct}) can be written as 
\begin{equation}
\label{cct}
6a\dot{a}^2=C a^{-3}+C_T,
\end{equation}
Now dividing equation (\ref{c}) by equation (\ref{cct}) we get
\begin{equation}
\frac{\sqrt{F(\phi)}d\phi}{da}=\pm
\frac{\sqrt{C}a^{-3}}{\sqrt{\frac{C}{6}a^{-4}+\frac{C_T}{6}a^{-1}}}.
\end{equation}
Taking only the positive part and integrating we find
\begin{equation}
\label{phia}
\phi(a)=\left\{
\frac{(m+2)}{3\lambda}\sinh^{-1}\left[\sqrt{\frac{C}{C_T}}a^{\frac{-3}{2}}\right]+ C_3 \right\}^{\frac{2}{m+2}},
\end{equation}
where $C_3$ is an integration constant. The field blows up at $a=0$ and takes a finite constant(the integration constant) value for large $a$. The blowing up of the field is the big bang singularity. The classical behaviour of the field as a function of the scale factor is shown in figure $1$.

\subsection{The Wheeler-DeWitt Equation}

Using the usual quantization procedure let us form the Schr\"{o}dinger-Wheeler-DeWitt equation (also known as Wheeler-DeWitt equation) for our super-Hamiltonian believing that the super-Hamiltonian operator annihilates the wave function. So with $P_a \rightarrow -i\partial_a$, $P_{\phi} \rightarrow -i\partial_{\phi}$, $P_T \rightarrow -i\partial_T$, $\hbar=1$ and

\begin{equation}
\label{h}
\hat{\cal H}\Psi(a,\phi,t)=0,
\end{equation}
we get from equation (\ref{sh1})
\begin{equation}
\frac{1}{24a}\frac{\partial^2 \Psi}{\partial a^2} - \frac{1}{4F(\phi)a^3}\frac{\partial^2 \Psi}{\partial \phi^2} - i\frac{\partial \Psi}{\partial t}=0.
\end{equation}
In order the Hamiltonian operator $\hat{\cal H}$ to be self-adjoint the inner product of any two wave functions $\Psi_1$ and $\Psi_2$ must take the form \cite{r26,r27}
\begin{equation}
\label{sa}
(\Psi_1,\Psi_2)=\int_{(a,\phi)} a\,\Psi_1^{\ast}\,\Psi_2 \,da\,d\phi ,
\end{equation}
and the restrictive boundary conditions being
\begin{equation}
\Psi(0,\phi,t)=0 \quad \mbox{or} \quad  \left. \frac{\partial \Psi(a,\phi,t)}{\partial a}\right|_{a=0}=0 \quad,
\end{equation}
or
\begin{equation}
\Psi(a,0,t)=0 \quad \mbox{or} \quad \left. \frac{\partial \Psi(a,\phi,t)}{\partial \phi}\right|_{\phi=0}=0.
\end{equation}
Now we have to apply the method of separation of variables. Putting 
\begin{equation}
\Psi(a,\phi,t)=e^{iEt}\eta(a,\phi),
\end{equation}
we get
\begin{equation}
a^2\frac{\partial^2 \eta}{\partial a^2} - \frac{6}{F(\phi)}\frac{\partial^2 \eta}{\partial \phi^2} + 24Ea^3\,\eta=0,
\end{equation}
where $E$ is the separation constant.
If we put 
\begin{equation}
\eta(a,\phi)=x(a)y(\phi),
\end{equation}
we will get
\begin{equation}
\label{ye}
\frac{\partial^2 y}{\partial \phi^2} + \frac{F(\phi)k^2}{24}y=0
\end{equation}
and
\begin{equation}
\label{xe}
a^2\frac{\partial^2 x}{\partial a^2} + 24Ea^3x + \frac{k^2}{4}x=0
\end{equation}
where $\frac{k}{2}$ is the separation constant. Putting the same expression for $F(\phi)$ as before
\begin{equation}
F(\phi)=6\lambda^2\phi^m \quad, \quad m\neq-2 \quad ,
\end{equation}
we get the solution of equation of equation (\ref{ye}) as
\begin{equation}
y(\phi)=k^{\frac{1}{m+2}}\sqrt{\phi}\left[c_1J_{\frac{1}{m+2}}\left(\frac{k\lambda}{m+2}\phi^{\frac{m+2}{2}}\right)+c_2J_{\frac{-1}{m+2}}\left(\frac{k\lambda}{m+2}\phi^{\frac{m+2}{2}}\right)\right].
\end{equation}
The solution of equation (\ref{xe}) is also known and is 
\begin{equation}
x(a)=\sqrt{a}\left[c_3J_{\nu}\left(\frac{2}{3}\sqrt{24E}a^{\frac{3}{2}}\right)+c_4J_{-\nu}\left(\frac{2}{3}\sqrt{24E}a^{\frac{3}{2}}\right)\right],
\end{equation}
where $\nu=\frac{\sqrt{1-k^2}}{3}$ and $c_{(1,2,3,4)}$ are integration constants.
If we consider $c_2=c_4=0$ the boundary conditions are well satisfied and the wave function of equation (\ref{h}) is
\begin{equation}
\label{psi}
\psi(a,\phi,t)=c_5\,\sqrt{a\phi}\,k^{\frac{1}{m+2}}\,J_{\nu}\left(\frac{2}{3}\sqrt{24E}a^{\frac{3}{2}}\right)\,J_{\frac{1}{m+2}}\left(\frac{k\lambda}{m+2}\phi^{\frac{m+2}{2}}\right)\,e^{iEt} \quad,
\end{equation}
where $c_5$ is the combination of preceding integration constants. If we want to construct the wave packet as a superposition of the eigenfunctions then then we get
\begin{equation}
\Psi_{wp}(a,\phi,t)=\sqrt{a\phi}\int_{q=0}^{\infty}\int_{k=0}^1 A(q)B(k)J_{\nu}\left(qa^{\frac{3}{2}}\right)J_{\frac{1}{m+2}}\left(\frac{k\lambda}{m+2}\phi^{\frac{m+2}{2}}\right)\,e^{i\frac{3}{32}q^2t}\,dq\,dk   \quad,
\end{equation}
correct upto a constant factor. Here $q=\frac{2}{3}\sqrt{24E}$ and $A(q)$ and $B(k)$ are suitable functions constructed from the integration constants so that the integration can be carried out swiftly. The factor $k^{\frac{1}{m+2}}$ of equation (\ref{psi}) is absorbed inside $B(k)$.
If we choose 
\begin{equation}
A(q)=q^{\nu+1}\,e^{-\gamma q^2},
\end{equation}
the integration on $q$ is known \cite{r52} and we get
\begin{equation}
\Psi_{wp}=\sqrt{a\phi}\frac{e^{-\frac{a^3}{4s}}}{(2s)}\int_{k=0}^1 B(k)\frac{a^{\frac{\sqrt{1-k^2}}{2}}}{{(2s)}^{\frac{\sqrt{1-k^2}}{3}}}J_{\frac{1}{m+2}}\left(\frac{k\lambda}{m+2}\phi^{\frac{m+2}{2}}\right)dk \quad,
\end{equation}
where $s=\gamma-i\frac{3}{32}t$. At this stage if we approximate $\sqrt{1-k^2} \sim 1$, then the general pattern of the resulting wave packet will remain same \cite{r18}. Now if we choose 
\begin{equation}
B(k)=k^{\frac{1}{m+2}+1}\sqrt{1-k^2} \quad,
\end{equation}
the integration can be done \cite{r53} and in final form 
\begin{equation}
\label{wpf}
\Psi_{wp}=a\phi^{\frac{-2m-3}{2}}\frac{e^{-\frac{a^3}{4s}}}{(2s)^{\frac{4}{3}}}J_{\frac{2m+5}{m+2}}\left(\frac{\lambda}{m+2}\phi^{\frac{m+2}{2}}\right)
\end{equation}
correct upto a constant factor. In the next section using this wave packet we will follow two different approaches to recover the classical solutions of the dynamical variables. In figure 2 we have plotted the square of the wave packet against the scale factor and the field $\phi$ for different values of time. There we see that the wave packet has two distinct peaks and both the peaks have the same height. This can be interpreted as that our universe might have evolved from one of these peaks and tunnelled with time. Looking at the figure we can also see that the peak height is getting reduced with time but progresses along increasing scale factor with a finite spread. Throughout this evolution we also notice that the field $\phi$ almost remained constant which is expected. As we have already approximated the super Hamiltonian to study the dynamics at a very early universe, so we can say that the field approached a nearly constant value within a very short time after big bang. 
\begin{figure}
\label{peaks}
\begin{tabular}{c}
\includegraphics[width=6.5cm,height=6cm]{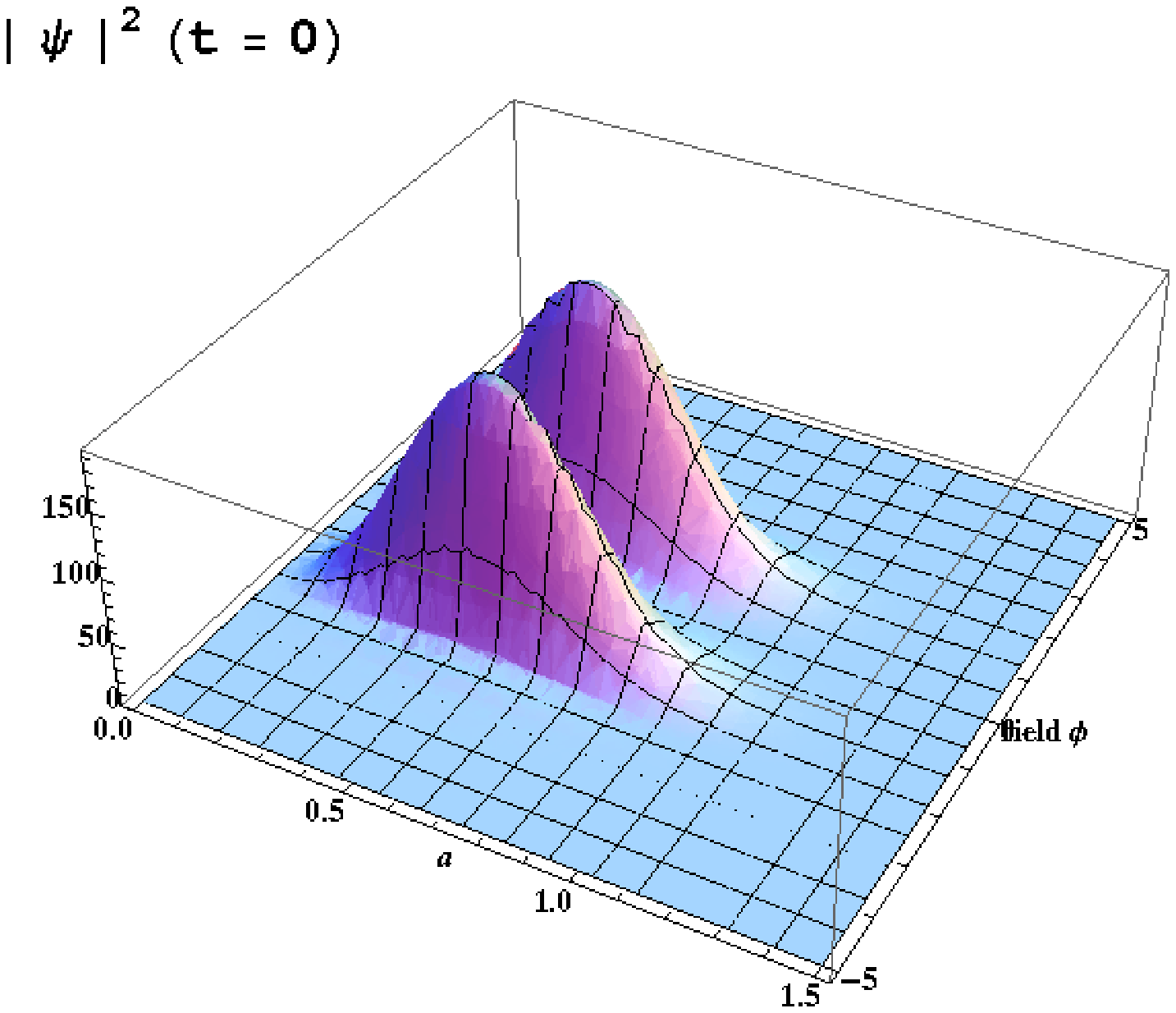} \hspace{1cm}
\includegraphics[width=6.5cm,height=6cm]{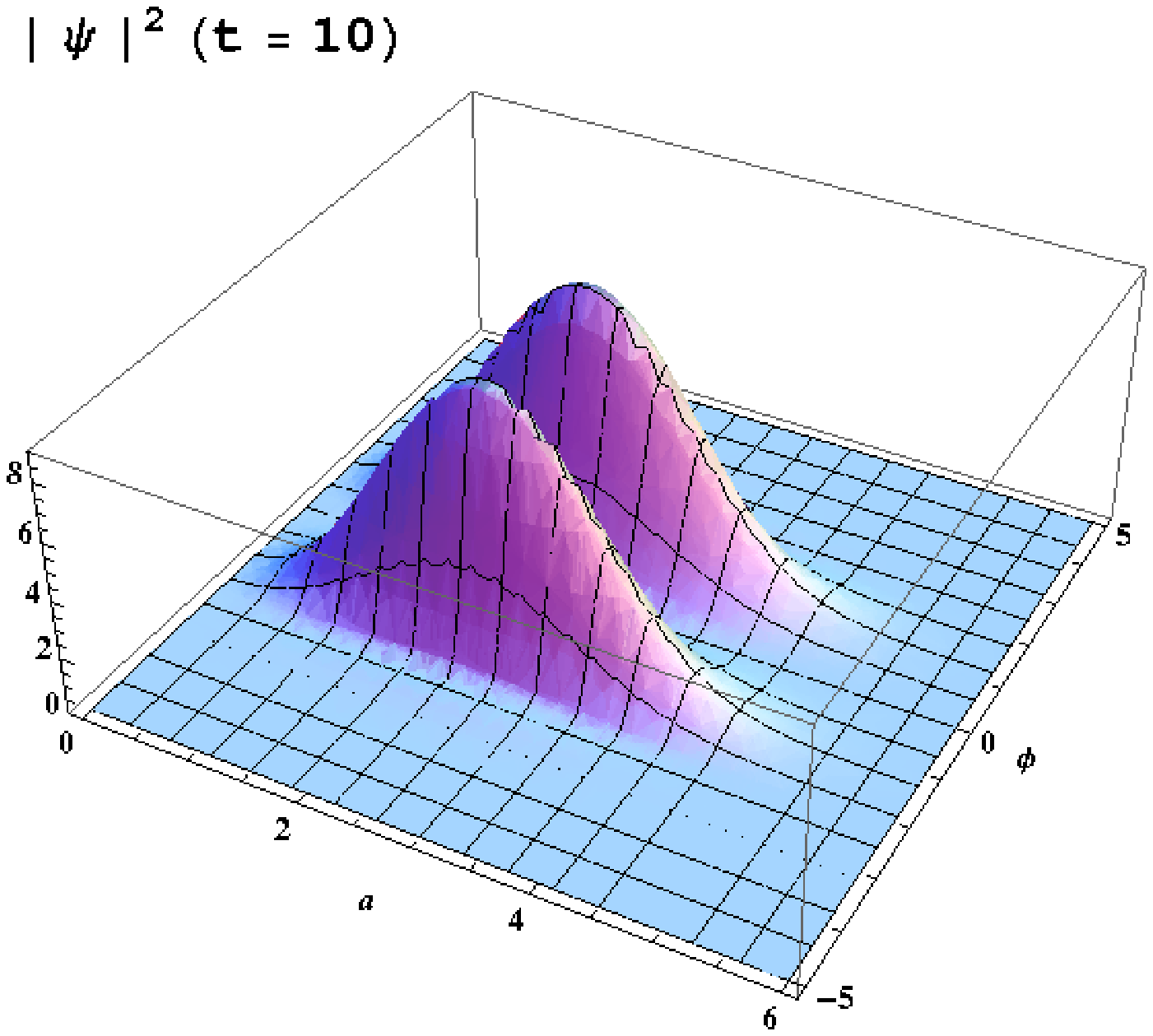}
\end{tabular}

\begin{tabular}{c}
\includegraphics[width=6.5cm,height=6cm]{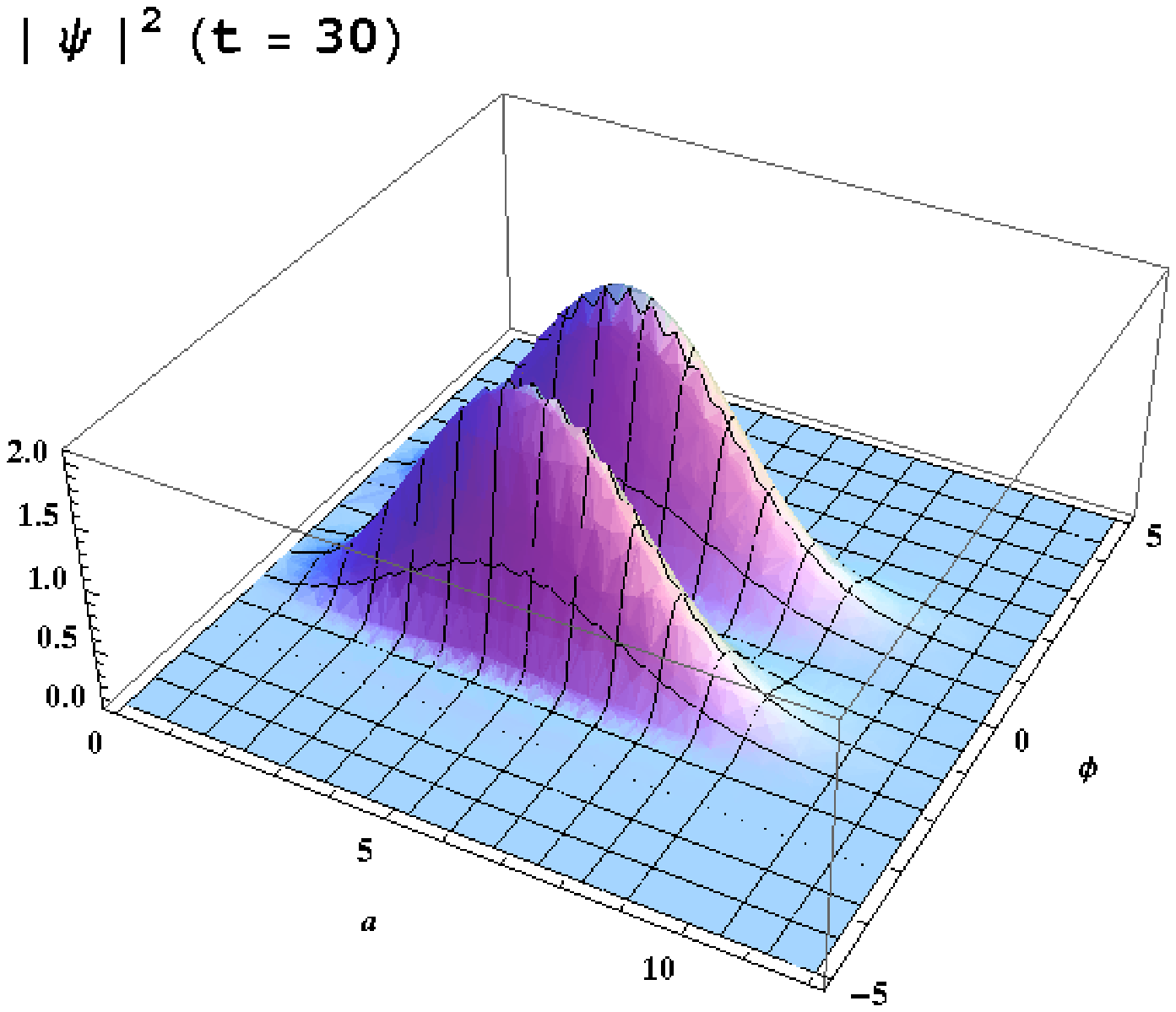} \hspace{1cm}
\includegraphics[width=6.5cm,height=6cm]{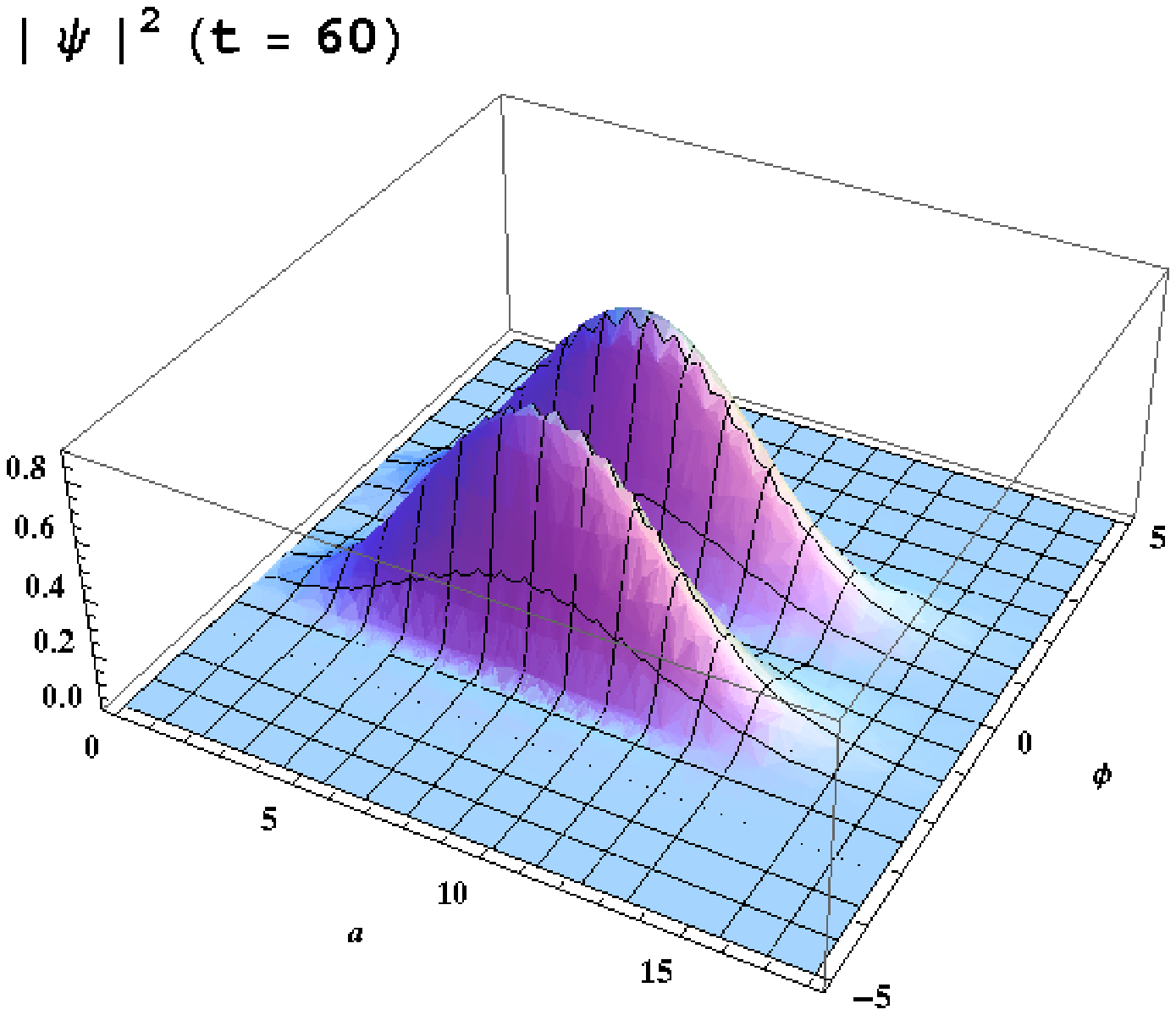}
\end{tabular}
\caption{\footnotesize  The behaviour of ${\vert\Psi_{wp}(a,\phi,t)\vert}^2$ at four different times $0,10,30,60$ is shown. The plots are with numerical values $m=2, \lambda=4,\gamma=0.09375$. With other admissible values of these parameters the same nature of the curves are repeated.}
\end{figure}

\subsection{Expectation values from the Many-World Interpretation of Quantum Mechanics}

Using the many-world interpretation of quantum mechanics \cite{r51} we can calculate the expectation value of the scale factor $a$ and the field $\phi$. Using equation (\ref{sa}) and equation (\ref{wpf}) one can calculate the expectation value of the scale factor
\begin{equation}
<a>(t)=\frac{\int_{a=0}^{\infty}\int_{\phi=-\infty}^{\infty}a\Psi^*a\Psi da d\phi}{\int_{a=0}^{\infty}\int_{\phi=-\infty}^{\infty}a\Psi^*\Psi da d\phi}.
\end{equation}
Straightforward calculation gives
\begin{equation}
\label{at}
<a>(t)=\frac{\Gamma(\frac{5}{3})}{\Gamma(\frac{4}{3})}\left[2\gamma +\frac{9}{512\gamma}t^2\right]^{\frac{1}{3}}.
\end{equation}
Similarly we can calculate the expectation value for the field $\phi$. In this case
\begin{equation}
<\phi>(t)=\frac{\int_{a=0}^{\infty}\int_{\phi=-\infty}^{\infty}a\Psi^* \phi \Psi da d\phi}{\int_{a=0}^{\infty}\int_{\phi=-\infty}^{\infty}a\Psi^*\Psi da d\phi} \quad,
\end{equation}
which on integration yields
\begin{equation}
\label{phit}
<\phi>(t)=\frac{\int_{\phi=-\infty}^{\infty}\phi^{\frac{-2m-1}{2}}\left[J_{\frac{2m+5}{m+2}}\left(\frac{\lambda}{m+2}\phi^{\frac{m+2}{2}}\right)\right]^2d\phi}{\int_{\phi=-\infty}^{\infty} \phi^{\frac{-2m-3}{2}}\left[J_{\frac{2m+5}{m+2}}\left(\frac{\lambda}{m+2}\phi^{\frac{m+2}{2}}\right)\right]^2d\phi}=\mbox{constant}.
\end{equation}
From equation (\ref{at}) we can see that the expectation value of the scale factor do not vanish at $t=0$. So it represents a bouncing non-singular universe and may behave like a dust dominated classical universe at later times. From equation (\ref{phit}) we see that the expectation value of $\phi$ is independent of time. It agrees well with the interpretation from equation (\ref{fit}) and with equation (\ref{phia}) (which says that $\phi$ catches a constant value for large values of time and scale factor) and the behaviour of ${\vert\Psi_{wp}\vert}^2$ as shown in figure $2$.

\subsection{Behaviour of the scale factor and the field from the Ontological Interpretation of quantum mechanics}

Let us now evaluate the expression for the scale factor $a$ and the field $\phi$ using ontological interpretation of quantum mechanics. In this formalism the wave function $\Psi$ is written in the form
\begin{equation}
\label{a1}
\Psi=\Omega e^{i\Theta}
\end{equation}
and the conjugate momenta corresponding to the variable $q$ is given by
\begin{equation}
\label{a2}
p_q=\frac{\partial \Theta}{\partial q}
\end{equation}
Portraying the wave function of equation (\ref{wpf}) in the form of equation (\ref{a1}) we find 
\begin{equation}
\label{a3}
\Theta(a,\phi,t)=\frac{4}{3}\arctan \left(\frac{3t}{32\gamma}\right) - \frac{3a^3t}{128\gamma^2 + \frac{9}{8}t^2}   \quad .
\end{equation}
So with equation (\ref{a2})
\begin{equation}
\label{a4}
p_a=-\frac{9a^2t}{128\gamma^2 + \frac{9}{8}t^2} \quad ,\quad p_{\phi}=0 \quad .
\end{equation}
From equation (\ref{aphi}) with our previously chosen gauge $N=1$ and with equation (\ref{a4}) we get
\begin{equation}
a(t)=a_0\left[\frac{512}{3}\gamma^2 + \frac{3}{2}t^2\right]^{\frac{1}{3}}
\end{equation}
and
\begin{equation}
\phi(t)=\mbox{constant} \quad .
\end{equation}
$a_0$ is an integration constant. Here also the scale factor is not zero for $t=0$, so represents a non-singular bouncing universe and $\phi$ is independent of time. So the many-world interpretation and the ontological interpretation give us the same behaviour for the scale factor and the field. But in general both these interpretations may not predict the same behaviour of the variables(seen in anisotropic cosmological models \cite{r22}).

\section{Conclusions}

Here we have studied a flat minisuperspace FRW cosmological model with Chaplygin gas and a scalar field coupled to the metric. Schutz's mechanism has allowed us to obtain the Schr\"{o}dinger-Wheeler-DeWitt (SWDW) equation for this minisuperspace in our early universe. The Chaplygin gas gave us the only remaining matter degree of freedom which finally played the role of cosmic time. We have constructed a well behaved wave packet from the linear superposition of the wave function of the SWDW equation. While solving the SWDW equation we did not bother about the factor ordering of the position and momentum operators present in the equation and it is seen that the behaviour of the constructed wave packet remains same for other factor orderings. Most interestingly we observe two peaks in the wave function, which could be interpreted as, that our universe might have evolved from one of those peaks(analogous to states in quantum mechanics). But the choice completely depends upon the initial conditions imposed on the field. Here we have observed that the wave packet propagates in the direction of the scale factor with time but the value of the field remained constant. We have calculated the time evolution of the expectation value of the scale factor and the field using the many-world interpretation of quantum mechanics and found that a bouncing and non-singular universe. The divergence of the field is also seen at big bang which assumes a constant value at later times. We have also calculated the time evolution of the scale factor and the field using ontological interpretation of quantum mechanics and found the same results.

\section*{Acknowledgements}
The author is very much thankful to Prof. Narayan Banerjee for helpful discussions and guidance.


\begin{thebibliography}{100}
\bibitem{r1} A.~Y.~Kamenshchik, U.~Moschella and V.~Pasquier, {\it Phys.~Lett.~B} {\bf 511}, 265 (2001) [arXiv:gr-qc/0103004].
\bibitem{r2} M.~C.~Bento, O.~Bertolami and A.~A.~Sen, {\it Phys.~Rev.~D} {\bf66}, 043507 (2002)  [arXiv:gr-qc/0202064].
\bibitem{r3} R.~Jackiw, [arXiv:physics/0010042].
\bibitem{r4} M.~C.~Bento, O.~Bertolami and A.~A.~Sen, {\it Phys.~Rev.~D} {\bf67},  063003 (2003)  [arXiv:astro-ph/0210468].
\bibitem{r5} R.~Bean and O.~Dore, {\it Phys.~Rev.~D} {\bf 68}, 023515 (2003) [arXiv:astro-ph/0301308].
\bibitem{r6} J.~C.~Fabris, S.~V.~Goncalves and P.~E.~de Souza, {\it Gen.~Rel.~Grav.}~ {\bf 34}, 53 (2002) [arXiv:gr-qc/0103083]
\bibitem{r7} R.~Colistete, J.~C.~Fabris, S.~V.~Goncalves and P.~E.~de Souza, [arXiv:gr-qc/0210079].
\bibitem{r8} N.~Ogawa, {\it Phys.~Rev.~D} {\bf62}, 085023 (2000)  [arXiv:hep-th/0003288].
\bibitem{r9} G.~M.~Kremer, {\it Gen.~Rel.~Grav.}~{\bf 35}, 1459 (2003) [arXiv:gr-qc/0303103].
\bibitem{r10} M.~R.~Setare, {\it Phys.Lett.~B} 644, 99 (2007).
\bibitem{r11} M.~Bouhmadi-L\'{o}pez, P.~V.~Moniz, Phys.~Rev.~D \textbf{71}, 063521 (2005).
\bibitem{r12} P.~Pedram, S.~Jalalzadeh and S.~S.~Gousheh, {\it Int.~J.~Theor.~Phys.} {\bf 46} :3201-3208 (2007).
\bibitem{r13} T.~Barreiro, A.A.~Sen, {\it Phys.~Rev.~D} \textbf{70}, 124013 (2004).
\bibitem{r14} M.~Heydari-Fard and H.~R.~Sepangi, to appear in {\it Phys.~Rev.~D}, [arXiv: 0710.2666].
\bibitem{r15} D. Saez and V.J. Ballester, {\it Phys. Lett.} A {\bf 113} (1986) 467
\bibitem{r16} J. Socorro, M. Sabido and A. Urena-Lopez, {\it Classical and quantum cosmology of the Saez-Ballester theory} (arXiv: 0904.0422 [gr-qc])
\bibitem{r17} J. Socorro, M. Sabido, M.A. Sanchez and M.G. Frias Palos, {\it REVISTA MEXICANA DE FI\'SICA} {\bf 56} (2) 166171 (2010).
\bibitem{r18} Babak Vakili, {\it Phys. Lett. B} {\bf 688} 129 (2010).
\bibitem{r19} B. F. Schutz, {\it Phys. Rev. D}{\bf 2}, 2762 (1970).
\bibitem{r20} B. F. Schutz, {\it Phys. Rev. D}{\bf 4}, 3559 (1971).
\bibitem{r21} F.G. Alvarenga, J.C. Fabris, N.A. Lemos and G.A. Monerat, {\it Gen.Rel.Grav.} 34 (2002) 651-663
\bibitem{r22} F.G. Alvarenga, A.B. Batista, J.C. Fabris and S.V.B. Gonsalves, {\it Gen.Rel.Grav.} 35 (2003) 1659-1677.
\bibitem{r23} P. Pedram and S. Jalalzadeh, {\it Phys.Lett. B} {\bf 659}:6-13,2008.
\bibitem{r24} V.~G.~Lapchinskii and V.~A.~Rubakov, {\it Theor.~Math.~Phys.}~{\bf 33}, 1076 (1977).
\bibitem{r25} R. Arnowitt, S. Deser and C. W. Misner, {\it Gravitation: An Introduction to Current Research}, edited by L. Witten, Wiley, New York (1962).
\bibitem{r26} F.~G.~Alvarenga and N.~A.~Lemos, {\it Gen.~Rel.~Grav.}~{\bf 30}, 681 (1998).
\bibitem{r27} N.~A.~Lemos, {\it J.~Math.~Phys.}~{\bf 37}, 1449 (1996).
\bibitem{r51} F.J. Tipler, {\it Phys. Rep.} {\bf 137} (1986) 231.
\bibitem{r52} W. W. Bell, {\it Special Functions for Scientists and Engineers}, D.Van Nostrand Company Inc., London, 1968.
\bibitem{r53} M. Abramowitz and I.A. Stegun, {\it Handbook of Mathematical Functions}, New York: Dover (1972).
\end{thebibliography}
\end{document}